\documentclass[12pt, a4paper]{article} % pdfを取り込めない。。。
\usepackage{amssymb, amsmath}
\usepackage{graphicx}
\usepackage{lmodern} % \normalfontがlatin modernになる。 本文のフォント（デフォルト）

\usepackage[margin=25mm]{geometry}

\usepackage[superscript]{cite}
\usepackage{color} % 赤追記
\usepackage{ulem} % 取り消し線 %\soutで取り消し線. \uwaveで下波線

\usepackage{caption}
\usepackage{textcomp}

\usepackage[compact]{titlesec}
\titleformat{\section}{\large\bfseries}{\thesection.}{15pt}{}
\titleformat{\subsection}{\itshape}{\thesubsection}{15pt}{}
\titleformat{\subsubsection}{\itshape}{\thesubsubsection}{15pt}{}

\makeatletter % overciteで使用。
\renewcommand{\@biblabel}[1]{#1)} % 参考文献を1)と表記 % overciteで使用。
\makeatother

\title{
\large{
\textbf{Cell--cell and Cell--noise Interactions of Bacterial Cells \\
in a Shallow Circular Pool \\
and Transitions of Collective Motions}
\vspace{-10pt} % maketitle内のspacingは末尾に記述する。
}
}
\author{
\normalsize{Ryojiro Honda, Sora Umeda, and Jun-ichi Wakita}}

\date{
\normalsize{\textit{Department of Physics, Chuo University, Bunkyo, Tokyo 112-8551, Japan}}
}

\begin{document}
\maketitle

%\begin{abstract}
%%\abst{
%\ \ We have experimentally investigated the transitions of collective motions of bacterial cells
%in a shallow circular pool using the bacterial species $Bacillus$ $subtilis$. 
%In our previous paper, 
%we reported that 
%the collective motions were classified into six phases on a phase diagram with two parameters,
%namely, the reduced cell length $\lambda$ and the cell density $\rho$.
%%
%%%% methods
%In this study, 
%we focused on the sharp transitions at $\lambda\cong0.1(\equiv\lambda_{\rm C1})$ with low values of $\rho$ 
%between the \textit{random motion} phase and the \textit{one-way rotational motion} phase.
%By introducing the order parameter $Q$, 
%which measures the aligned cell motion along the circumferential direction of a pool,
%the transitions at $\lambda=\lambda_{\rm C1}$ were clearly characterized.
%On the basis of the detailed observations of single-cell trajectories in a pool,
%we verified that the effect of cell--noise interactions was widely distributed.
%We conclude that, even in such random environment, 
%the sharp transitions of collective motions were caused by the cell--cell interactions at $\lambda=\lambda_{\rm C1}$.
%%}
%\end{abstract}

% 2020 arxiv
\vspace{-20pt}
We have experimentally investigated the transitions of collective motions of bacterial cells
in a shallow circular pool using the bacterial species $Bacillus$ $subtilis$. 
In our previous paper, 
we reported that 
the collective motions were classified into six phases on a phase diagram with two parameters,
namely, the reduced cell length $\lambda$ and the cell density $\rho$.
%
%%% methods
In this study, 
we focused on the sharp transitions at $\lambda\cong0.1(\equiv\lambda_{\rm C1})$ with low values of $\rho$ 
between the \textit{random motion} phase and the \textit{one-way rotational motion} phase.
By introducing the order parameter $Q$, 
which measures the aligned cell motion along the circumferential direction of a pool,
the transitions at $\lambda=\lambda_{\rm C1}$ were clearly characterized.
On the basis of the detailed observations of single-cell trajectories in a pool,
we verified that the effect of cell--noise interactions was widely distributed.
We conclude that, even in such random environment, 
the sharp transitions of collective motions were caused by the cell--cell interactions at $\lambda=\lambda_{\rm C1}$.

%
%\begin{document}

%\maketitle

%青字の修正候補を赤字で記しました。赤字のみの部分は追記候補になります。

%また、青字破線でコメントを記しました。

%\section{}
%\input{abst_v2.tex} %20181122
%\input{abst_v3_20181205.tex} %20181205
%\input{abst_v4_20181206.tex} %20181206, katori check
%\input{abst_v5_20181210.tex} %20181210
%\input{abst_v6_20181211.tex} %20181211
%\input{abst_v7_20181212.tex} %20181212
%\input{abst_v8_20181213.tex} %20181213

%\newpage

%%%%%%%%%%%%%%%%%%%%%%%%%%%%%%%%%%%%%%%%%%%%%%%%%%% section1
%%%%%%%%%%%%%%%%%%%%%%%%%%%%%%%%%%%%%%%%%%%%%%%%%%
\vspace{15pt}
\section{Introduction}
%\input{sect1_v1.tex}
%\input{sect1_v3_20181009.tex} 
%\input{sect1_v4} %
%\input{sect1_v5.tex} % 20181120
%\input{sect1_v6_20181205.tex}
%\input{sect1_v7_20181206.tex} % 20181206, katori
%\input{sect1_v8_20181210.tex} % 20181210
%\input{sect1_v9_20181211.tex} % 20181211
%\input{sect1_v10_20181212.tex} % 20181212, katori
%\input{sect1_v10_20181212_2.tex} % 20181212_2, katori, pm
%\input{sect1_v11_20181213_1.tex} % 20181213_1, am
%\input{sect1_v12_20181213_2.tex} % 20181213_2, pm
%\input{sect1_v13_20181215.tex} % 20181215
%\input{sect1_v14_20181216.tex} % 20181216
%\input{sect1_v15_20181217.tex} % 20181217
%\input{sect1_v16_20190128.tex} % 20190128
%\input{sect1_v17_20190206.tex} % 20190206
%\input{sect1_v18_20190209.tex} % 20190209
%\input{sect1_v19_20190214.tex} % 20190214
%\input{sect1_v20_20190216.tex} % 20190216
%\input{sect1_v21_20190222.tex} % 20190222
%\input{sect1_v22_20190318.tex} % 20190318
%%%%%%%%%%%%%%%%%%%%%%%%%%%%%%%%%%%%%%%%%%%%%%%%%%
\hspace{15pt} Various types of ordered behavior including the construction of vortexlike structures 
were observed in flocks of birds, schools of fish, marches of social insects, and migrations of bacteria.
% 引用
%\cite{VZ2012, BCC2008, CCG2010, HHR2010, BVD2006, CBC1996, WRI1998, SA2012, WDH2012}.
\cite{VZ2012, BCC2008, CCG2010, HHR2010, BVD2006, CBC1996, WRI1998, SA2012, WDH2012}$^{)}$
It was also reported that 
elongating bacteria produce complicated structures in folding and filling processes on an agar surface 
by cell multiplications.
%\cite{HWK2015, MMR2016}
\cite{HWK2015, MMR2016}$^{)}$

Migrating bacteria in a growing colony show a variety of collective behavior 
depending on environmental conditions,
although a bacterial colony is one of the simplest biological systems.
% 引用
%\cite{MHK2004, NSW1996, HWK2005, TKM2009}.
\cite{MHK2004, NSW1996, HWK2005, TKM2009}$^{)}$
Vicsek {\it et} {\it al}. proposed a statistical mechanics model 
called the {\it self-}{\it propelled} {\it particle} (SPP) model 
to demonstrate such collective motions.
% 引用
%\cite{Vicsek2001, VCB1995, CSV1997, HGG2008, BVJ2012}.
\cite{Vicsek2001, VCB1995, CSV1997, HGG2008, BVJ2012}$^{)}$
Their model shows the dynamical phase transition 
such that the ordered motions with aligned directions 
suddenly emerge from chaotic motions in the self-propelled particle system 
at a critical noise amplitude or at a critical particle density.
The origin of the transition is considered to be the competition between the effect that 
each particle tends to be aligned in the averaged direction of its neighboring particles 
and 
the individual fluctuations of moving directions due to noises.
The hydrodynamic motions of cells including the creation and annihilation of vortices 
have also been universally observed 
in growing colonies of several bacterial species.
%\cite{CBC1996,WRI1998}.
\cite{CBC1996,WRI1998}$^{)}$
Wioland $et$ $al$. reported that 
cells in a highly concentrated bacterial droplet form a single stable vortex 
that can rotate in both clockwise and counterclockwise directions in the droplet.
% 引用
%\cite{WWD2013}.
\cite{WWD2013}$^{)}$
By comparing their experimental results with a model, 
they argued that the global confinement condition
and the curvature of droplet boundaries play important roles in realizing 
such a steady state having a single vortex in the system. 
Similar self-organized structures involving vortices and clusters have also been reported 
in other confinement systems.
% 引用
%\cite{WWD2016, BIG2017, NAS2018, WH2017, BCD2015, DBG2018}.
\cite{WWD2016, BIG2017, NAS2018, WH2017, BCD2015, DBG2018}$^{)}$
% NRM2007, 
In an experimental study, however, it is generally difficult to verify 
which is essential for the emergence of collective motions 
among cell--cell interactions, cell--boundary interactions, 
and individual fluctuations due to noises.

\begin{figure}[ht]
\centering
\includegraphics[clip, width=0.5\hsize]{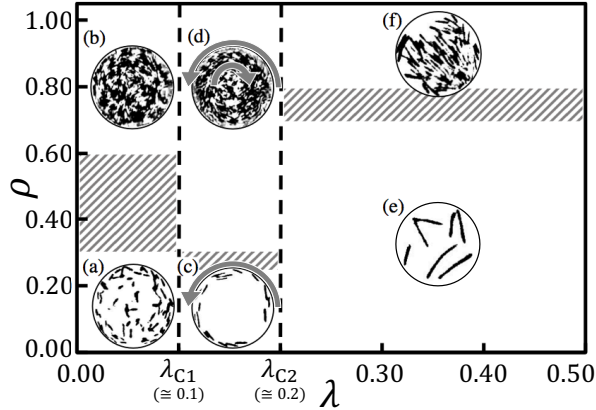} % for_arXiv
%\vspace{-10pt}
\caption{
%Figure caption.
Phase diagram of collective motions of bacterial cells in a shallow circular pool 
with two parameters $\lambda$ and $\rho$:
(a) \textit{random motion} phase, (b) \textit{turbulent motion} phase, 
(c) \textit{one-way rotational motion} phase, (d) \textit{two-way rotational motion} phase, 
(e) \textit{random oscillatory motion} phase, and (f) \textit{ordered oscillatory motion} phase.
The rotational directions in (c) and (d) are shown by the arrows.
The two vertical dashed lines show
the critical values $\lambda_{\rm C1}\cong0.1$ and $\lambda_{\rm C2}\cong0.2$, 
at which the collective motions are changed considerably.
}
%\label{figlabel}
\label{diagram}
\end{figure}

In our previous study, 
we experimentally investigated the collective motions of bacterial cells in a shallow circular pool 
using the bacterial species $Bacillus$ ($B$.) $subtilis$.
% 引用
%\cite{WTY2015}.
\cite{WTY2015}$^{)}$
The pools were made on the surface of an agar plate, 
in which the collective motions of rod-shaped bacterial cells were observed.
We introduced two relevant parameters, 
namely, the reduced cell length $\lambda$, 
which was defined as the ratio of the averaged cell length to the pool diameter,
and the average cell density $\rho$.
As shown by the phase diagram (Fig. \ref{diagram}), 
the observed collective motions have been classified into six dynamical phases:
(a) \textit{random motion}, 
(b) \textit{turbulent motion}, 
(c) \textit{one-way rotational motion}, 
(d) \textit{two-way rotational motion},
(e) \textit{random oscillatory motion}, and
(f) \textit{ordered oscillatory motion}.
In the \textit{random motion} phase, bacterial cells move independently and randomly in a pool.
In the \textit{turbulent motion} phase, bacterial cells form groups 
showing hydrodynamic motions with the creation and annihilation of vortices.
In the \textit{one-way rotational motion} phase, 
bacterial cells move counterclockwise along the brim of a pool 
keeping their axes parallel with the brim.
This suggests that the brim of a pool acts as a nearly slip boundary for a bacterial cell. 
Note that in Ref.\ \citen{WWD2013}, 
however, the interface of a droplet acts as a nearly no-slip boundary.
In the \textit{two-way rotational motion} phase, 
bacterial cells move counterclockwise in the outer region of a pool and clockwise in the inner region of a pool.
In the \textit{random oscillatory motion} phase, 
bacterial cells move forward and backward repeatedly in their axial directions.
In this phase, the axial oscillatory motions of bacterial cells seem to be individually random and uncorrelated.
In the \textit{ordered oscillatory motion} phase, 
on the other hand,
if we observe the system for a sufficiently long time,
intermittent ordering of the axial directions of bacterial cells is observed.
As indicated in Fig.\ \ref{diagram} by two dashed lines perpendicular to the $\lambda$ axis, 
there are two critical values of $\lambda$, 
$\lambda_{\rm C1}\cong0.1$ and $\lambda_{\rm C2}\cong0.2$.
By crossing these two lines in the phase diagram,
marked changes in collective motions occur.

In this study, 
we focused on the transitions at $\lambda=\lambda_{\rm C1}$ at the low values of $\rho$
between the $random$ $motion$ phase
and the $one$-$way$ $rotational$ $motion$ phase.
The transitions at $\lambda=\lambda_{\rm C1}$ were characterized 
by introducing the order parameter $Q$, 
which measures the aligned cell motion in the circumferential direction of a pool.
If the reduced cell length $\lambda$ increases with a constant cell density $\rho$, 
then more collisions between cells occur and the alignment effect of neighboring cells is enhanced.
In this sense, $\lambda$ is a parameter representing the coupling strength of cell--cell interactions.

%\textcolor{red}{At the vicinity of the boundary of the pool in these phases, 
%the counterclockwise motion of bacterial cells along the brim of the pool were observed.
%This behavior was seemed to be induced by interactions 
%between a bacterial cell and the brim (boundary) of the pool,
%so that we regarded the effect as the cell-boundary interactions.}

%In order to study cell-boundary interactions 
%and individual fluctuation of bacterial motions,
To study cell--cell interactions,
we have examined the systems in which only a single bacterial cell is moving in a pool.
As a matter of course, 
no cell--cell interactions exist in such single cell systems.
%\textcolor{red}{
The parameter $\tilde{Q}$ for 
the \textit{one-way rotational motion} in the single cell systems,
%}
which was defined 
%\textcolor{red}{
as the time-averaged velocity component parallel to the brim of a pool,
%}
did not show any systematic dependence on $\lambda$.
We found instead that 
$\tilde{Q}$ approximately linearly depended on the time-average speed $v_{\rm ave}$ of a bacterial cell.
We verified that 
$v_{\rm ave}$ was determined by the frequency of the change in the moving direction in each trajectory of a cell.
The change in the moving direction of a cell will be attributed to 
%\textcolor{red}{
the cell--noise interactions between a cell and irregularities on the surface of a shallow pool,
%}
%irregularities 
%on the bottom and the brim of a shallow pool (the surface of an agar plate) 
which cannot be controlled in our experimental setting.

Combining the experimental results of the original systems with a large number of bacterial cells 
and the single cell systems,
we conclude that the transitions at $\lambda=\lambda_{\rm C1}$ 
from the \textit{random motion} phase to the \textit{one-way rotational motion} phase
were caused by the cell--cell interactions in the systems involving uncontrolled cell--noise interactions.

All through the experiments, we used the wild-type strain OG-01 of $B.$ $subtilis$.
%\cite{MHK2004}.
\cite{MHK2004}$^{)}$
$B.$ $subtilis$ cells are rod-shaped with peritrichous flagella 
and swim straightforwardly in water by bundling and rotating their flagella.
When a small number of cells are inoculated on the surface of semisolid agar plates, 
they form a colony by cell motility and multiplication.
Growing colonies present five different patterns depending on two environmental parameters 
$C_{\rm a}$ (agar concentration) and $C_{\rm n}$ (nutrient concentration).
%\cite{MHK2004}.
\cite{MHK2004}$^{)}$
In particular, when $C_{\rm a}$ is intermediate (7 g/L $<C_{\rm a}<$ 8.5 g/L) 
and $C_{\rm n}$ is high ($C_{\rm n}>$ 10 g/L),
the interface of a growing colony repeatedly advances (in the migration phase) 
and rests (in the consolidation phase),
and finally yields a concentric-ring pattern. 
The bacterial cells at the growing front of a concentric-ring pattern 
have been observed to repeat elongation and contraction, synchronizing with the periodic colony growth.
%
%\cite{MHK2004, WSI2001}.
\cite{MHK2004, WSI2001}$^{)}$

Shallow circular pools are made on the surface of an agar plate
by scattering glass beads on the growing front of a concentric-ring pattern 
and then removing the glass beads.
%
%\cite{WTY2015}.
\cite{WTY2015}$^{)}$
Bacterial cells trapped in a pool show a variety of collective motions.
We can control $\lambda$ by changing the timing of making pools on the growing front,
since the cell length of bacteria differs at different times in the growing front
in the concentric-ring pattern formation.
Thanks to the local fluctuation of the cell density at the growing front, 
a variety of pools with different cell densities $\rho$ can be made.

This paper is organized as follows. 
In Sect. 2, we explain the experimental procedures.
Experimental results are given in Sect. 3.
In Sect. 4, we discuss our experimental results and future problems.

%\clearpage

%%%%%%%%%%%%%%%%%%%%%%%%%%%%%%%%%%%%%%%%%%%%%%%%%%

%\clearpage

%%%%%%%%%%%%%%%%%%%%%%%%%%%%%%%%%%%%%%%%%%%%%%%%%%
% section2
%%%%%%%%%%%%%%%%%%%%%%%%%%%%%%%%%%%%%%%%%%%%%%%%%%
\section{Experimental Procedure}
%\input{sect2_v2.tex}
%\input{sect2_v3.tex}
%\input{sect2_v4.tex} % 20181120
%\input{sect2_v5_20181204.tex} % 20181204
%\input{sect2_v6_20181206.tex} % 20181206, katori
%\input{sect2_v7_20181210.tex} % 20181210
%\input{sect2_v8_20181212.tex} % 20181212
%\input{sect2_v9_20181213.tex} % 20181213
%\input{sect2_v10_20190128.tex} % 20190128
%\input{sect2_v11_20190209.tex} % 20190209
%\input{sect2_v12_20190318.tex} % 20190318
%%%%%%%%%%%%%%%%%%%%%%%%%%%%%%%%%%%%%%%%%%%%%%%%%%%
\hspace{15pt}
First, we prepared semisolid agar plates by following the procedures explained below.
5 g of sodium chloride (NaCl), 
5 g of dipotassium hydrogen phosphate (K$_2$HPO$_4$),
and 30 g of Bacto-Peptone (Becton, Dickinson and Co.) as nutrient 
were dissolved in 1 L of distilled water.
The environmental parameter $C_{\rm n}$ was set to 30 g/L by adjusting the concentration of Bacto-Peptone.
Then, the solution was adjusted to pH 7.1 by adding 6N hydrochloric acid (HCl). 
Furthermore,  the solution was mixed with 8.3 g of Bacto-Agar (Becton, Dickinson and Co.), 
which determines the softness of the semisolid agar plates.
The environmental parameter $C_{\rm a}$ was set to 8.3 g/L 
by adjusting the concentration of Bacto-Agar.
The environmental condition given by these $C_{\rm a}$ and $C_{\rm n}$ values
generates a typical concentric-ring pattern of $B$. $subtilis$ colonies.
The solution was autoclaved at 121 $^\circ$C for 15 min, 
and 20 ml of the solution was poured into each sterilized plastic petri dish of 88 mm inner diameter.
The thickness of the semisolid agar plates was about 3.2 mm.
After solidification at room temperature for 60 min, the semisolid agar plates were dried at 50 $^\circ$C for 90 min.

Next, we prepared a bacterial suspension with an optical density of 0.5 at a wavelength of 600 nm.
We inoculated 3 \textmu l of the bacterial suspension on the surface of each semisolid agar plate.
The optical density of 0.5 corresponds to a bacterial density of about 10$^4$ cells per \textmu l.
The semisolid agar plates were left at room temperature for about 60 min to dry the bacterial suspension droplet.

Then, we incubated the semisolid agar plates in a humidified box at 35 $^\circ$C and 90\% RH.
Bacterial cells at the inoculation spot 
grew and multiplied by cell division without migration during the lag phase period of about 7 h.
After the lag phase, the first migration started and two-dimensional colony expansion was observed.
About 2 h later, they stopped migrating and entered the first consolidation phase.
They did not move but underwent cell division actively for about 5 h. 
Afterwards, they exhibited the migration phase and the consolidation phase alternately.
Thus, we obtained a concentric-ring pattern of the bacterial colonies.

After that, we scattered glass beads of 50 $\pm$ 2 \textmu m diameter (Unitika, SPM-50)
in the vicinity of the growing front of a concentric-ring pattern
at the start of the third migration phase or the third consolidation phase of colony growth.
After we removed the beads from the agar surface using adhesive tape, 
circular pools were made on the surface.
Bacterial cells under a bead were trapped and were moving in a pool.
%\cite{WTY2015}.
\cite{WTY2015}$^{)}$
The pool depth was about 1 \textmu m,
which was much smaller than the pool diameter.
The thickness of each bacterial cell was about 0.5 \textmu m,
which was approximately the same scale of the pool depth.
As a result, two-dimensional motions of bacterial cells were realized in the pools.
%No bacterial cells coming in or getting out from the pool were observed during the observation periods, 
%so that the number of bacterial cells was unchanged.
During an observation period, 
which was set to be 10 min, corresponding to the half of a cell division cycle, 
the cell length was almost constant.
We have confirmed that
there was no inflow of bacterial cells into a pool and no outflow of bacterial cells from a pool.
Therefore, the cell density $\rho$ was almost constant 
and the collective motions of bacterial cells were stationary during an observation period.
Furthermore, water was always supplied to a pool from the agar surface,
so that we were able to observe the motions of bacterial cells for relatively long time periods, 
which were typically more than 10 min.

We observed and video-recorded
the collective motions of bacterial cells at 30 Hz with a high-speed microscope (Keyence, VW-9000)
linked to an optical microscope (Nikon, DIAPHOT-TMD).
We defined the area of a circular pool as the area covered by the trajectories traced by bacterial cells 
in the time duration of 10 min. 
The pool diameter was calculated from the area of a pool.
In the above analysis, we used an image and motion analysis library (OpenCV).
On the other hand, we measured cell lengths in a snapshot captured from the video by hand.
The cell density in each pool was given by dividing the total area of bacterial cells 
in a pool in a snapshot by the area of a pool.
To acquire the velocity field of bacterial motion in the \textit{random motion} phase and the \textit{one-way rotational motion} phase, 
we used particle image velocimetry (PIV) software (Library, Flow-PIV). 
%%%%%%%%%%%%%%%%%%%%%%%%%%%%%%%%%%%%%%%%%%%%%%%%%%%

%\newpage % 改ページ

%%%%%%%%%%%%%%%%%%%%%%%%%%%%%%%%%%%%%%%%%%%%%%%%%%
% section3
%%%%%%%%%%%%%%%%%%%%%%%%%%%%%%%%%%%%%%%%%%%%%%%%%%
\section{Experimental Results}
%\input{sect3_v1.tex}
%\input{sect3_v3.tex}
%\input{sect3_v4.tex} % 20181120
%\input{sect3_v6_20181205.tex} % 20181203
%\input{sect3_v7_20181206.tex} % 20181206
%\input{sect3_v8_20181210.tex} % 20181210
%\input{sect3_v9_20181211.tex} % 20181211
%\input{sect3_v10_20181212.tex} % 20181212
%\input{sect3_v11_20181213.tex} % 20181213
%\input{sect3_v12_20181213_2.tex} % 20181213_2, pm
%\input{sect3_v13_20181215.tex} % 20181215
%\input{sect3_v14_20181216.tex} % 20181216
%\input{sect3_v15_20181217.tex} % 20181217
%\input{sect3_v16_20190128.tex} % 20190128
%\input{sect3_v17_20190206.tex} % 20190206
%\input{sect3_v18_20190209.tex} % 20190209
%\input{sect3_v19_20190214.tex} % 20190214
%\input{sect3_v20_20190216.tex} % 20190216
%\input{sect3_v21_20190222.tex} % 20190222
%\input{sect3_v22_20190223.tex} % 20190223
%\input{sect3_v23_20190318.tex} % 20190318
%%%%%%%%%%%%%%%%%%%%%%%%%%%%%%%%%%%%%%%%%%%%%%%%%%%
\subsection{Transition between random motion phase and one-way rotational motion phase}
\subsubsection{Introducing order parameter $Q$}

\hspace{15pt}
To clarify the differences between the cell motions in the \textit{random motion} phase 
and in the \textit{one-way rotational motion} phase,
we focused on the circumferential direction component of the cell velocity in a pool
as explained in the following.

\begin{figure}[htbp]
\centering
\includegraphics[clip,width=1.0\hsize]{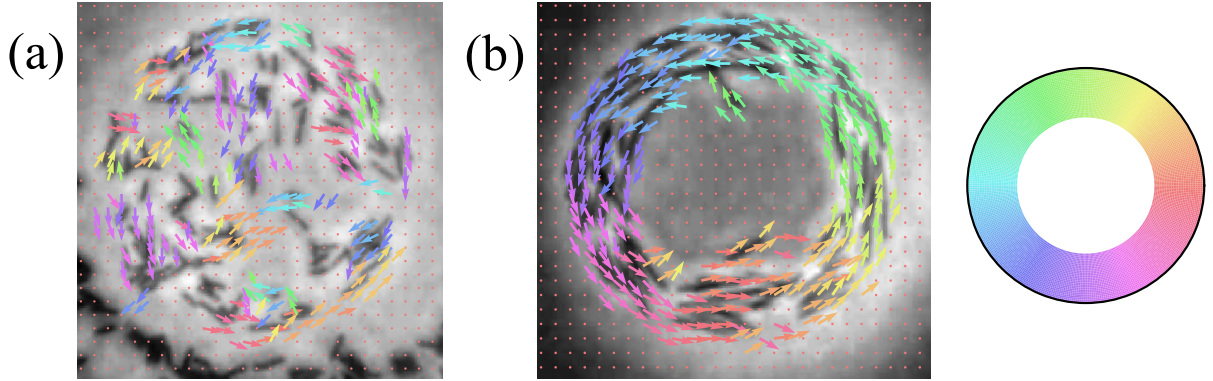} % for_arXiv(article環境)
%\includegraphics[clip,width=0.5\hsize]{figs/CM_results_3_1_1.pdf}
%\vspace{-10pt}
\caption{
%Figure caption.
(Color online)
Typical velocity fields for the \textit{random motion} phase at $\lambda=0.07$ in (a)
and for the \textit{one-way rotational motion} phase at $\lambda=0.17$ in (b).
Velocity fields overlaid on the snapshots of bacterial cells, 
in which the orientations of velocity vectors are indicated 
by arrows with colors assigned as shown by the color ring.
}
%\label{figlabel}
\label{fig:CM_typical_Q_R}
\end{figure}

We performed time-series measurements at intervals of 1/30 s of the bacterial velocity fields 
in the range of $0.05<\lambda<0.20$.
One observation period is 60 s, 
so that each time-series measurement consists of 1,800 discrete velocity fields.
The typical velocity fields of cell motions in the \textit{random motion} phase 
and in the \textit{one-way rotational motion} phase
are shown in the snapshots
in Figs.\ \ref{fig:CM_typical_Q_R}(a) and \ref{fig:CM_typical_Q_R}(b), respectively.

We put the origin of two-dimensional coordinates at the center of a circular pool 
and introduce the polar radius $r$ and the polar angle $\theta$.
To analyze the velocity fields, 
we consider a lattice of a square mesh of 1.88 \textmu m spacing
on a circular pool and use the polar coordinates ($r$, $\theta$) to specify the lattice points.
At each lattice point ($r$, $\theta$), 
{\boldmath $\hat{S}$}($r$, $\theta$) is defined as a unit tangent vector in the circumferential direction of the pool,
where the positive direction is assumed to be in the counterclockwise direction.

At each time $t$, 
the velocity vectors {\boldmath $V$}($t$; $r$, $\theta$) were measured 
and their normalized vectors {\boldmath $\hat{V}$}($t$; $r$, $\theta$) were obtained by
{\boldmath$V$}($t$; $r$, $\theta$)/$|${\boldmath$V$}($t$; $r$, $\theta$)$|$
at all lattice points ($r$, $\theta$).
Then we defined the local order parameter $q(t; r, \theta)$ by the inner product

%\begin{equation}
%q(t; r, \theta) = {\bf \hat{V}}(t; r, \theta) \cdot {\bf \hat{S}}(r, \theta)
%\end{equation}

\begin{equation}
q(t; r, \theta) = \boldsymbol{\hat{V}}(t; r, \theta) \cdot \boldsymbol{\hat{S}}(r, \theta)
\end{equation}

\noindent
at each time $t$ and at each lattice point ($r$, $\theta$).
Then for each $r$, we averaged $q(t; r, \theta)$ over the observation time-period $T=60$ s 
(by summing over 1,800 discrete velocity fields at different times)
and averaged it over polar angles $\theta$ $\in$ [0, 2$\pi$).
That is, we obtained 
a discretized approximation on the lattice for the quantity

%\begin{equation}
%q(r) = \frac{1}{T} \int_0^T dt \int_0^{2\pi} d\theta\ q(t; r, \theta).
%\end{equation}

%\begin{equation}
%q(r) = \frac{1}{T} \sum_{t=0}^T \sum_{\theta=0}^{2\pi} q(t;r, \theta).
%\end{equation}

\begin{equation}
q(r) = \frac{1}{T} \int_0^T dt\ \frac{1}{2\pi} \int_0^{2\pi} d\theta\ q(t; r, \theta).
\end{equation}

\noindent
Since the radius $a$ of a pool was distributed,
%\cite{WTY2015}, 
\cite{WTY2015}$^{)}$
we introduced the normalized radius coordinate as $R=r/a$.
The $R$-dependent order parameter was then defined as

\begin{equation}
Q(R) = q(aR).
\end{equation}

\noindent
By the definition mentioned above,
if all cells are undergoing counterclockwise \textit{one-way rotational motion} at any place in a pool,
we have $Q(R)=1$ for each $R\in(0, 1)$, 
while if cell motions are completely random,
$Q(R)=0$ for each $R\in(0, 1)$.
Hence, $Q(R)$ will act as the order parameter
for the \textit{one-way rotational motion}.

Figure\ \ref{fig:CM_relation_Q_lambda}(a) shows the profiles of $Q$($R$) 
for $\lambda=0.07<\lambda_{\rm C1}$ and $\lambda=0.17>\lambda_{\rm C1}$, 
where $\lambda_{\rm C1}\cong0.1$.
The profiles have a maximum value at $R\simeq0.8$,
while $Q(R)\simeq0$ for 
%\textcolor{red}{
$R<0.5$.
%}.
From now on, 
we focus on the maximum value, 
and define the order parameter $Q$
for the \textit{one-way rotational motion}
as 

\begin{equation}
Q=\max_R\ Q(R).
\end{equation}

\subsubsection{Behavior of order parameter $Q$}
\label{sect:CM_Q}

\begin{figure}[htbp]
\centering
\includegraphics[clip,width=0.5\hsize]{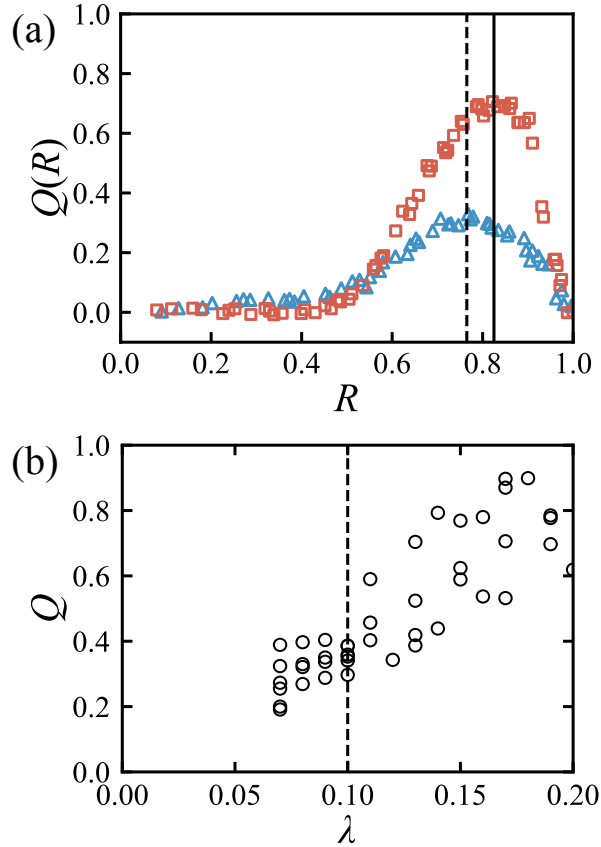} % for arXiv
%\vspace{-10pt}
\caption{
%Figure caption.
(Color online)
Measurement of the order parameter $Q$.
(a) Profiles of $Q$($R$) for $\lambda=0.07<\lambda_{\rm C1}$ (blue triangles) 
and for $\lambda=0.17>\lambda_{\rm C1}$ (red squares),
where $\lambda_{\rm C1}\cong0.1$.
Both profiles have maximum values at $R\simeq0.8$,
while they have $Q(R)\simeq0$ for 
%\textcolor{red}{
$R<0.5$.
%}.
(b) Values of $Q$ plotted against $\lambda$ for $0.11\leq\rho\leq0.30$.
For $\lambda\leq\lambda_{\rm C1}\cong0.10$,
$Q$ remains at small values $\simeq0.3$,
while $Q$ evidently increases with $\lambda$ for $\lambda\geq\lambda_{\rm C1}$.
}
%\label{figlabel}
\label{fig:CM_relation_Q_lambda}
\end{figure}

\hspace{15pt}
Figure\ \ref{fig:CM_relation_Q_lambda}(b) shows the $\lambda$-dependence of the order parameter $Q$
in the vicinity of $\lambda_{\rm C1}$.
There, we plotted $Q$ only for the range of $0.11\leq\rho\leq0.30$ for the following reasons.
The upper limit $\rho\simeq0.30$ is the threshold value between 
the \textit{one-way rotational motion} phase and the \textit{two-way rotational motion} phase
as shown in Fig.\ \ref{diagram}.
In the \textit{one-way rotational motion} phase, 
bacterial cells were often observed near the brim of the pool, 
so that the local cell density near the brim was effectively higher than
the density $\rho$ averaged over the whole pool.
When $\rho\geq 0.11$, 
bacterial cells filled the outer region of the pool
and the \textit{one-way rotational motion} became stable.
As shown in Fig.\ \ref{fig:CM_relation_Q_lambda}(b),
$Q$ remains at small values $\simeq0.3$ for $\lambda\leq\lambda_{\rm C1}\cong0.1$,
while it is evident that $Q$ increases with $\lambda$ for $\lambda\geq\lambda_{\rm C1}$.
The quantity $Q=Q(\lambda)$ indeed acts as the order parameter for the \textit{one-way rotational motion}.
(Strictly speaking, 
the order parameter for \textit{one-way rotational motion} 
should be zero for \textit{random motion}.
A possible improvement of the definition of the order parameter is proposed in Sect.\ 4.)

%\clearpage

%\subsubsection{Relation between $\lambda$ and $v_{\rm ave}$}
%
%Figure \ref{fig:CM_speed_property}(a) shows a typical speed histogram 
%at $\lambda=0.15$ and $\rho=0.14$.
%It was made from bacterial speeds at 1,800 lattice points in all local bacterial velocity fields.
%The histogram exhibited a gaussian distribution, 
%so that the spatio-temporal average speed for 600 seconds givenby 34.8 \textmu m/s corresponded to the peak of the gaussian distribution.
%for 600 sec given by $v_{\rm ave}=34.8$ \textmu m/s.
%Figure \ref{fig:CM_speed_property}(b) shows $\lambda$ dependence of $v_{\rm ave}$.
%Most of $v_{\rm ave}$ were distributed in the vicinity of 30 \textmu m/s.
%Consequently, $v_{\rm ave}$ was almost constant independently of $\lambda$.

%
% Results for Collective Motion 1
%

%
% Results for Collective Motion 2
%

%
% Results for Collective Motion 3
%
%\begin{figure}[htbp]
%\centering
%%\includegraphics[clip,width=.5\hsize]{figure.eps}
%%\includegraphics[clip,width=.5\hsize]{figs/apef_2018.png}
%%\includegraphics[clip,width=1.0\hsize]{figs/CM_results_3_1_3.pdf}
%\includegraphics[clip,width=0.5\hsize]{figs/CM_results_3_1_3.pdf}
%%\vspace{-10pt}
%\caption{
%%Figure caption.
%Relation between $\lambda$ and $v_{\rm ave}$.
%(a) A typical speed histogram at $\lambda=0.15$ and $\rho=0.14$. 
%The vertical dotted line indicates the spatio-temporal average speed $v_{\rm ave}$ for 60 seconds.
%(b) $\lambda$ dependence of $v_{\rm ave}$.
%Most of $v_{\rm ave}$ were distributed in the vicinity of 30 ($\mu$m/s).
%}
%%\label{figlabel}
%\label{fig:CM_speed_property}
%\end{figure}

%%%%%%%%%%%%%%%%%%%%%%%%%%%%%%%%%%%%%%%%%%%%%%%%%%%%%
%%%%%%%%%%%%%%%% sect 3.2   single bacterial system %%%%%%%%%%%%%%%%%%%%
%%%%%%%%%%%%%%%%%%%%%%%%%%%%%%%%%%%%%%%%%%%%%%%%%%%%%

%\newpage

%\clearpage

%\subsection{\textcolor{red}{Cell motions in single cell systems}}
%\subsubsection{\textcolor{red}{Introducing parameter $\tilde{Q}$}}
%\subsection{\textcolor{red}{Effects of cell--cell interactions and fluctuation of bacterial motions}}
\subsection{Effects of cell--cell interactions and fluctuation of bacterial motions}
\subsubsection{Characterization of bacterial motions in single cell systems}

\hspace{15pt}
To examine 
the effects of 
%\textcolor{red}{
cell--cell interactions
%} 
on the transitions between
the \textit{random motion} phase and the \textit{one-way rotational motion} phase,
we have examined single cell systems
with only one bacterial cell moving in each pool.
In the original systems with many bacterial cells in each pool,
the cell--cell interactions will be effective.
They may cause the tendency of the bacterial motions 
to make the individual direction of motion 
be aligned with the averaged moving direction of the neighboring cells,
and also the repulsive effects between bacterial cells.
As a matter of course, 
no cell--cell interactions exist in the single cell systems.

\begin{figure}[htbp]
\centering
\includegraphics[clip,width=0.5\hsize]{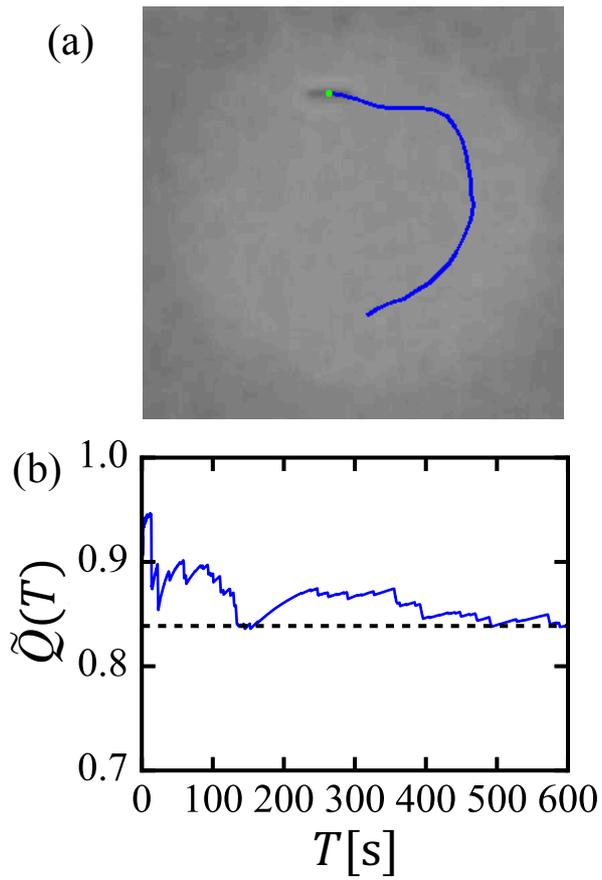} % for arXiv
%\vspace{-10pt}
\caption{
%Figure caption.
(Color online)
Tracking a bacterial cell in a single cell system.
(a) Trajectory of the geometric center of a bacterial cell for 1 s.
(b) Time average of the inner products $\tilde{Q}(T)$ defined by Eq.\ (\ref{eqn:definition_of_tilde_Q})
shown as a function of averaging time $T$.
}
%\label{figlabel}
\label{fig:ES_typical_Q}
\end{figure}

First, we tracked a bacterial cell in each single cell system 
at intervals of 1/30 s for the time duration of 600 s. 
Again, we used the two-dimensional polar coordinate system with the origin set at the center of the pool.
For each trajectory of the geometric center of a cell,
we obtained a time series of 18,000 polar coordinates ($r(t)$, $\theta(t)$).
Figure \ref{fig:ES_typical_Q}(a) shows a typical trajectory for 1 s.

Then, we evaluated the time-series velocity vectors {\boldmath$v$}($r(t)$, $\theta(t)$) 
from the differences in successive polar coordinates ($r(t)$, $\theta(t)$). 
We calculated the inner products of the normalized velocity vectors {\boldmath$\hat{v}$}($r(t)$, $\theta(t)$)
defined as {\boldmath$v$}($r(t)$, $\theta(t)$)/$|${\boldmath$v$}($r(t)$, $\theta(t)$)$|$
and the unit tangent vectors {\boldmath$\hat{s}$}($r(t)$, $\theta(t)$) defined as the unit vectors in the circumferential direction of a pool at the polar coordinates ($r(t)$, $\theta(t)$).
The positive direction of {\boldmath $\hat{s}$}($r(t)$, $\theta(t)$) is assumed to be 
in the counterclockwise direction.
We examined the behavior of the time average of the inner products defined as 

%\begin{equation}
%\tilde{Q}(t)=\frac{1}{t}\sum_{\tau=0}^{t} \hat{v}(r(\tau), \theta(\tau)) \cdot \hat{s}(r(\tau), \theta(\tau)),
%\end{equation}
\begin{equation}
\tilde{Q}(T)=\frac{1}{T}\sum_{t=0}^{T} \boldsymbol{\hat{v}}(r(t), \theta(t)) \cdot \boldsymbol{\hat{s}}(r(t), \theta(t))
\label{eqn:definition_of_tilde_Q}
\end{equation}

\noindent
by changing the averaging time duration $T$.
Figure \ref{fig:ES_typical_Q}(b) shows $\tilde{Q}(T)$
as a function of $T$.
This asymptotically approached the value for $T=600$ s, 
which is shown by the dotted line in Fig. \ref{fig:ES_typical_Q}(b).
By this consideration, 
we defined the parameter $\tilde{Q}$
for the \textit{one-way rotational motion} in the single cell systems
by $\tilde{Q}(600)$; $\tilde{Q}=\tilde{Q}(600)$.
%\textcolor{red}{
If a bacterial cell moves counterclockwise in the circumferential direction of a pool, 
we have $\tilde{Q}=1$, 
whereas if a bacterial cell moves completely randomly, $\tilde{Q}=0$.
%}

%%%%%%%%%%%%%%%%%%%%%%%%%%%%%%%%%%%%%%%%%%%%%%%%
%%% subsubsec 3.2.2
%%%%%%%%%%%%%%%%%%%%%%%%%%%%%%%%%%%%%%%%%%%%%%%%
\subsubsection{Behavior of parameter $\tilde{Q}$}

\hspace{15pt}
Figure \ref{fig:ES_Q_properties}(a) shows plots of the parameter $\tilde{Q}$ against $\lambda$
in the range of $0<\lambda<0.20$.
Note that $\tilde{Q}$ is always positive, 
which implies that bacterial cells tend to move counterclockwise in the circumferential direction of their pool.
%\textcolor{red}{
In the microscopic observation, 
when bacterial cells moved counterclockwise along the brim of a pool,
the axes of the cells were parallel to the brim.
On the other hand, 
when bacterial cells sometimes moved clockwise, 
the axes were not parallel to the brim and kept a certain angle against the brim.
Bacterial cells seemed unable to swim clockwise smoothly along the brim.
%}
%\textcolor{red}{
The difference in the axial directions to the brim 
between counterclockwise motion and clockwise motion
is due to 
some biological characteristics of the present species of bacterial cells,
hence we regard the interactions between a cell and the brim as
the cell--boundary interactions.
%} 
The values of $\tilde{Q}$ were widely distributed between 0 and 1, 
and seemed to have no systematic dependence on $\lambda$.
The critical value $\lambda_{\rm C1}$ found in Fig.\ \ref{fig:CM_relation_Q_lambda}(b) for the collective motions 
becomes meaningless in the single cell systems.

Then, we studied the average of bacterial cell speeds $v$($=|${\boldmath$v$}($r(t)$, $\theta(t)$)$|$) 
over the time duration of 600 s, which was denoted as $v_{\rm ave}$.
Figure \ref{fig:ES_Q_properties}(b) shows the plots of the parameter $\tilde{Q}$ 
against the time-average speed $v_{\rm ave}$. 
$\tilde{Q}$ seemed to increase in proportion to $v_{\rm ave}$.
%\textcolor{red}{
%The coordinates ($v_{\rm ave}$ [\textmu m/s], $\tilde{Q}$) 
%of the colored symbols shown as $\square$ (blue), $\triangle$ (green), 
%and $\Diamond$ (red) in Fig. \ref{fig:ES_Q_properties}(b) 
%are (20.5, 0.18), (35.2, 0.58), 
%and (51.1, 0.84), respectively.
The distributions of cell speed $v$ and the trajectories of a bacterial cell 
%for these plots 
for the colored plots in Fig. \ref{fig:ES_Q_properties}(b)
are shown in 
Fig. \ref{fig:ES_speed_hist_and_trajectory} in the next section.
%}

%%%%%%%%%%%%%%%%%%%%%%%%%%%%%%%%%%%%%%%%%%%%%%%%
%%% subsubsec 3.2.3
%%%%%%%%%%%%%%%%%%%%%%%%%%%%%%%%%%%%%%%%%%%%%%%%
\subsubsection{Distributions of cell speed $v$}
\label{sect:SC_distribution_v}

%\textcolor{red}{
%In Sect. 3.2.2, we found that $\tilde{Q}$ depended on $v_{\rm ave}$, 
%having no dependence on $\lambda$.
%Then, what does the difference of the time-average speed $v_{\rm ave}$ mean?
%In order to examine the property of $v_{\rm ave}$,
%}

\begin{figure}[htbp]
\centering
\includegraphics[clip,width=0.5\hsize]{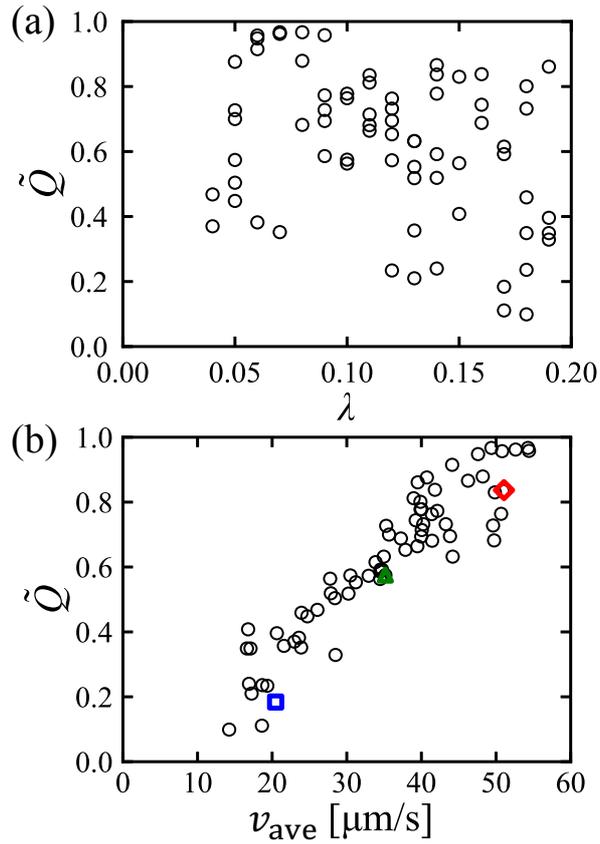} % for arXiv
%\vspace{-10pt}
\caption{
%Figure caption.
(Color online)
Behavior of the parameter $\tilde{Q}$ for the \textit{one-way rotational motion} in the single cell systems.
(a) Parameter $\tilde{Q}$ plotted against $\lambda$ ($0<\lambda<0.2$).
%\textcolor{red}{
$\tilde{Q}$ has no systematic dependence on $\lambda$.
%}
(b) Parameter $\tilde{Q}$ plotted against $v_{\rm ave}$.
%\textcolor{red}{
$\tilde{Q}$ seems to increase in proportion to $v_{\rm ave}$.
%}
%\textcolor{red}{
The coordinates ($v_{\rm ave}$ [\textmu m/s], $Q$) 
of the colored symbols shown as $\square$ (blue), $\triangle$ (green), 
and $\Diamond$ (red) are
(20.5, 0.18), (35.2, 0.58), 
and (51.1, 0.84), respectively.
The distributions of cell speed $v$ and the trajectories of a bacterial cell for these plots are shown in 
Fig. \ref{fig:ES_speed_hist_and_trajectory}.
%}
}
%\label{figlabel}
\label{fig:ES_Q_properties}
\end{figure}

\begin{figure}[htbp]
\centering
\includegraphics[clip,width=0.5\hsize]{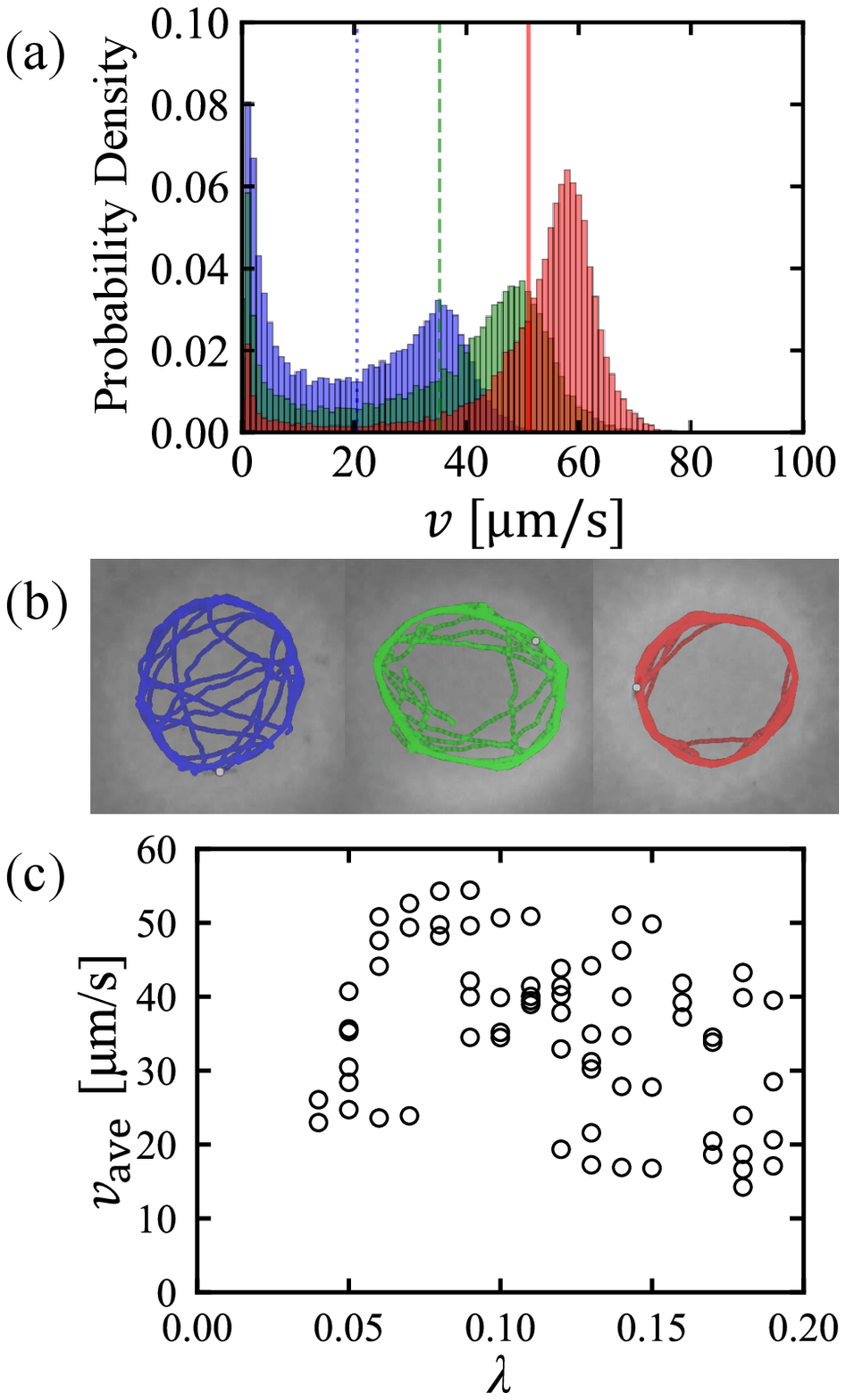}
%\vspace{-10pt}
\caption{
%Figure caption.
(Color online)
Comparison of the distributions of cell speed $v$.
(a) Probability densities of cell speed $v$ distributed in time 
shown for the different 
%\textcolor{red}{
time-average speeds,
$v_{\rm ave}=20.5$ \textmu m/s (blue dotted line), 
35.2 \textmu m/s (green dashed line), 
and 51.1 \textmu m/s (red solid line).
%}
%\textcolor{red}{
These values of $v_{\rm ave}$ correspond to the colored symbols
shown as $\square$ (blue), $\triangle$ (green), 
and $\Diamond$ (red)
in Fig. \ref{fig:ES_Q_properties}(b), respectively.
%}
%\textcolor{red}{
Each distribution was made from 18,000 successive speeds $v$.
%}
%These values correspond to the colored circles in Fig.\ \ref{fig:ES_Q_properties}(b).
(b) Trajectories of a bacterial cell for the different values of $v_{\rm ave}$ in (a) shown for 60 s.
The left, middle, and right pictures correspond to the distributions of $v$ 
for $v_{\rm ave}=20.5$, 35.2, and 51.1 \textmu m/s, respectively.
(c) Values of $v_{\rm ave}$ plotted against $\lambda$.
%\textcolor{red}{
$v_{\rm ave}$ did not seem to have any systematic dependence on $\lambda$.
%}
% left     : 20170718_06_480s
% center: 20170713_07_300s
% right   : 20170714_21_120s
}
%\label{figlabel}
\label{fig:ES_speed_hist_and_trajectory}
\end{figure}

\hspace{15pt}
In order to know what determines the time-average speed $v_{\rm ave}$
of a bacterial cell in the single cell systems, 
we compared the distributions of cell speed $v$ for different values of $v_{\rm ave}$.
Figure \ref{fig:ES_speed_hist_and_trajectory}(a) shows the histograms of the cell speed $v$ 
for different values of $v_{\rm ave}$.
Each histogram was made from 18,000 successive cell speeds $v$.
The bin range was set from 0 to 100 \textmu m/s with a bin width of 1 \textmu m/s.
The total area of the histograms was normalized to one, 
so that they give probability densities.
The three vertical lines indicate the values of $v_{\rm ave}$ for each histogram, 
which are 20.5 \textmu m/s (dotted line), 35.2 \textmu m/s (dashed line), 
and 51.1 \textmu m/s (solid line) from the left.
These values of $v_{\rm ave}$ were also shown by the colored symbols in Fig.\ \ref{fig:ES_Q_properties}(b).
The probability densities of $v$ had two peaks. 
One of them was at $v\cong0$ \textmu m/s and the other one was at $v>30$ \textmu m/s.
%\textcolor{red}{$v>30$} \textmu m/s.
%\textcolor{red}{
When the probability density at $v\cong0$ \textmu m/s was low, 
that at $v>30$ \textmu m/s was high and then $v_{\rm ave}$ was large.
When the probability density at $v\cong0$ \textmu m/s was high, 
that at $v>30$ \textmu m/s was low and then $v_{\rm ave}$ was small.
These tendencies suggest that $v_{\rm ave}$ strongly depends on the relationship between the two peaks.
%}
The trajectories of a bacterial cell for the different values of $v_{\rm ave}$ in 
Fig.\ \ref{fig:ES_speed_hist_and_trajectory}(a) are shown for 60 s in 
Fig.\ \ref{fig:ES_speed_hist_and_trajectory}(b).
Namely, the left, middle, and right pictures in Fig. \ref{fig:ES_speed_hist_and_trajectory}(b) 
correspond to the distributions of $v$ for $v_{\rm ave}=20.5$ , 35.2, and 51.1 \textmu m/s 
in Fig.\ \ref{fig:ES_speed_hist_and_trajectory}(a), respectively.
%\textcolor{red}{As $v_{\rm ave}$ increased}, 
As $v_{\rm ave}$ increased, 
the trajectories seemed to become less complicated, 
%\textcolor{red}{that is, the fluctuation of the bacterial motion decreased.}
that is, the fluctuation of the bacterial motion decreased.

Here, we examined the dependence of $v_{\rm ave}$ on $\lambda$.
Figure \ref{fig:ES_speed_hist_and_trajectory}(c) shows that the values of $v_{\rm ave}$ were widely distributed between 10 and 60 \textmu m/s, and did not seem to have any systematic dependence on $\lambda$.
%\textcolor{red}{
%This result supports the above expectation that 
%$v_{\rm ave}$ is determined by the behavior of the two peaks.
This result is consistent with the results shown in 
Figs.\ \ref{fig:ES_Q_properties}(a) and \ref{fig:ES_Q_properties}(b)
and with the possibility that $v_{\rm ave}$ depends on the relationship between the two peaks.
%}

%\clearpage

%
% Results for Elementary Substance 1
%

%
% Results for Elementary Substance 2
%

%
% Results for Elementary Substance 3
%
%%%%%%%%%%%%%%%%%%%%%%%%%%%%%%%%%%%%%%%%%%%%%%%%%%%

%\clearpage % 未出力の図を出力して改ページ

%%%%%%%%%%%%%%%%%%%%%%%%%%%%%%%%%%%%%%%%%%%%%%%%%%
% section4
%%%%%%%%%%%%%%%%%%%%%%%%%%%%%%%%%%%%%%%%%%%%%%%%%%
%\section{Discussion}
%\section{\textcolor{red}{Discussion and Future Problems}}
\section{Discussion and Future Problems}
%\input{sect4_v2_h1.tex}
%\input{sect4_v4_20181205} % 20181203
%\input{sect4_v6_20181210} % 20181206
%\input{sect4_v7_20181211} % 20181211
%\input{sect4_v8_20181213} % 20181213
%\input{sect4_v9_20181215} % 20181215
%\input{sect4_v10_20181216} % 20181216
%\input{sect4_v11_20181217} % 20181217
%\input{sect4_v12_20190128} % 20190128
%\input{sect4_v13_20190204} % 20190204
%\input{sect4_v14_20190209} % 20190204
%\input{sect4_v15_20190214} % 20190214
%\input{sect4_v16_20190216} % 20190216
%\input{sect4_v17_20190222} % 20190222
%\input{sect4_v18_20190318} % 20190318
%%%%%%%%%%%%%%%%%%%%%%%%%%%%%%%%%%%%%%%%%%%%%%%%%%

\begin{figure}[htbp]
\centering
\includegraphics[clip,width=0.5\hsize]{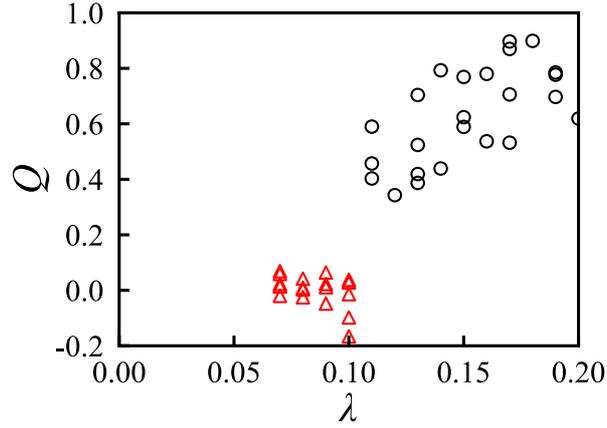} % for arXiv
%\vspace{-10pt}
\caption{
%Figure caption.
(Color online)
Dependence of the redefined order parameter $Q$ on $\lambda$. 
}
%\label{figlabel}
\label{fig:Discussion_Qt_R03}
\end{figure}

\hspace{15pt}
In Sect.\ \ref{sect:CM_Q},
we studied the transitions of the collective motions of bacterial cells in a shallow circular pool 
between the \textit{random motion} phase and the \textit{one-way rotational motion} phase
when the reduced cell length $\lambda$
was changed in the vicinity of $\lambda_{\rm C1}$.
% PARAGRAPH 2
From the dependence of $Q$ on $\lambda$ shown in Fig.\ \ref{fig:CM_relation_Q_lambda}(b),
a change in the behavior of $Q$ was recognized at $\lambda=\lambda_{\rm C1}$.
In the microscopic observation for $\lambda\geq\lambda_{\rm C1}$ 
(in the \textit{one-way rotational motion} phase),
we found that the bacterial cells were localized in the outer region of their pool 
and were moving counterclockwise along the brim of the pool.
%[The outer region is defined as the range of $0.6\leq R\leq1.0$ from the profile of $Q$($R$) 
%in Fig.\ \ref{fig:CM_relation_Q_lambda}(a).] 
(The outer region is defined as the range of $0.6\leq R\leq1.0$ from the profile of $Q$($R$) 
in Fig.\ \ref{fig:CM_relation_Q_lambda}(a).) 
As shown in Fig.\ \ref{fig:CM_relation_Q_lambda}(a), 
the value of $R$ at which $Q(R)$ attained its maximum was about 0.8.
Therefore, the order parameter $Q$ defined by $\max_R Q(R)$
indicates a typical value of $Q(R)$
for bacterial cell motions.
On the other hand, 
from the microscopic observation, 
we found that 
also in the systems with $\lambda\leq\lambda_{\rm C1}$ (in the \textit{random motion} phase), 
the bacterial cells in the outer region of the pool tended to move counterclockwise along the brim of the pool.
This tendency is considered to be due to the cell--boundary interactions.
As a result,
the values of $Q$ were in the range of $0.2\leq Q\leq0.4$
even for $\lambda\leq\lambda_{\rm C1}$ as shown in Fig.\ \ref{fig:CM_relation_Q_lambda}(a).
We considered, however, 
that the typical behavior of bacterial cell motions for $\lambda\leq\lambda_{\rm C1}$
should be measured in the inner region of a pool 
apart from the cell--boundary interactions.
Thus, 
here we redefine the order parameter $Q$ 
%\textcolor{red}{for $\lambda\leq\lambda_{\rm C1}$}
for $\lambda\leq\lambda_{\rm C1}$
as the value of $Q(R)$ at $R=0.3$ 
(the middle position of the inner region).
Figure\ \ref{fig:Discussion_Qt_R03} shows the dependence of $Q$ on $\lambda$, 
in which the values of $Q$ became almost zero for $\lambda\leq\lambda_{\rm C1}$.

The behavior of $Q$ reminds us of the dynamical phase transition in the SPP model.
%\cite{VZ2012, Vicsek2001, VCB1995}.
\cite{VZ2012, Vicsek2001, VCB1995}$^{)}$
In the SPP model, 
the transition of the collective motion of self-propelled particles is caused by the competitive relationship
between the effect that 
the individual moving direction is aligned with the averaged moving direction of its neighboring particles 
and the fluctuations of the moving directions due to noises.
There, for a specified particle, neighboring particles are defined as the particles 
in a circle of some given radius (interaction radius) centered at the position of the specified particle.
The transition occurs at the critical noise amplitude or at the critical particle density in the SPP model.
On the other hand, 
in our experiment, 
the transitions were observed at the critical value of the reduced cell length $\lambda_{\rm C1}$.
From  the microscopic observation, 
we saw that each rod-shaped bacterial cell had the tendency 
to be aligned its moving direction with the neighboring cells by the effect of cell--cell interactions.
This tendency seemed to 
be monotonically increasing as 
the length of bacterial cells increased.
Therefore, $\lambda$ is considered to correspond to the interaction radius in the SPP model.
%The existence of the critical interaction radius has to be verified in SPP model.

% PARAGRAPH 3
%\textcolor{red}{
To discuss the effect of cell--cell interactions,
we examined the single cell systems 
by defining the parameter $\tilde{Q}$ 
for the \textit{one-way rotational motion} in the single cell systems.
%}
Figure\ \ref{fig:ES_Q_properties}(a) shows the plots of the parameter $\tilde{Q}$ against $\lambda$.
%\textcolor{red}{
The values of $\tilde{Q}$ were always positive and were independent of $\lambda$.
This is due to the effect of cell--boundary interactions.
%}
No critical value $\lambda_{\rm C1}$ was found in the plots of $\tilde{Q}$
in the single cell systems having no cell--cell interactions.
This result indicates 
that the transitions between the \textit{random motion} phase and the \textit{one-way rotational motion} phase
are caused by the cell--cell interactions.

\begin{figure}[htbp]
\centering
\includegraphics[clip,width=0.5\hsize]{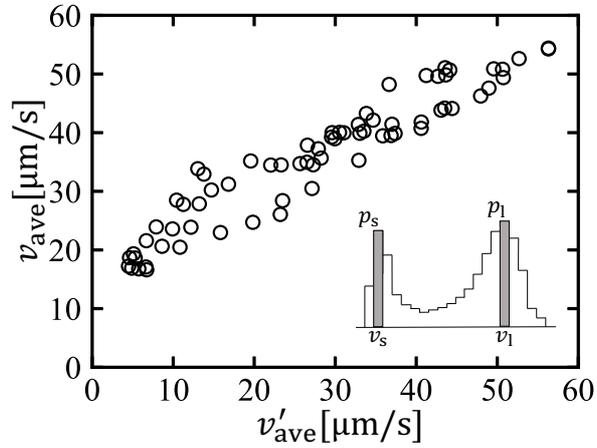} % for arXiv
%\vspace{-10pt}
\caption{
%Figure caption.
%\textcolor{red}{
%(Color online)
Relationship between $v^{\prime}_{\rm ave}$ and $v_{\rm ave}$.
% and the peaks of the distributions in single cell systems.
%(a) The schematic figure of coarse graining of distribution.
%The smaller value of $v$ is denoted as $v_{\rm s}$
%and the probability is represented as $p_{\rm s}$.
%The larger value of $v$ and the probability are shown as $v_{\rm l}$ and $p_{\rm l}$, respectively.
%Then $v^{\prime}$ is defined as Eq.\ (\ref{eqn:vprime}).
%(b) $v^{\prime}$ shows the tendency to monotonically decrease as $p_{\rm s}$ increases.
%The red solid curve indicates the exponential function defined as Eq.\ (\ref{eqn:exponential_function_vprime}) 
%with the fitting parameters obtained as Eq.\ (\ref{eqn:fitting_parameters_vprime}).
%(c) The relation between $v^{\prime}$ and $v_{\rm ave}$.
%They seem to be have a linear relationship.
$v_{\rm ave}$ seemed to be linearly dependent on $v^{\prime}_{\rm ave}$.
The inset shows a schematic figure of the distribution of cell speeds $v$ in a single cell system.
We focused on the two peaks at $v_{\rm s}$ ($\cong0$ \textmu m/s) and $v_{\rm l}$ ($>30$ \textmu m/s).
$p_{\rm s}$ and $p_{\rm l}$ are the probability densities at $v=v_{\rm s}$ and $v=v_{\rm l}$, respectively.
%} %end textcolor
}
%\label{figlabel}
\label{fig:Discussion_vprime_vave}
\end{figure}

\begin{figure}[htbp]
\centering
\includegraphics[clip,width=0.5\hsize]{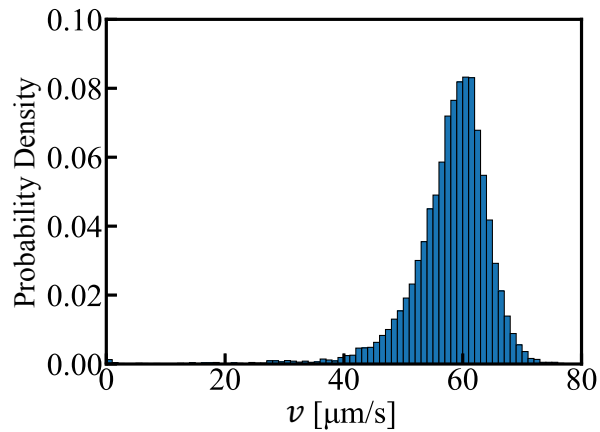} % for arXiv
%\vspace{-10pt}
\caption{
%Figure caption.
(Color online)
One-peak histogram for the distribution of cell speeds $v$ when $v_{\rm ave}=54.3$ \textmu m/s. 
%\textcolor{red}{
The histogram was made from 18,000 successive speeds $v$.
%}
}
%\label{figlabel}
\label{fig:Discussion_one_peak_histogram}
\end{figure}

% PARAGRAPH 4
In Sect.\ \ref{sect:SC_distribution_v}, 
we examined the dependence of $\tilde{Q}$ on the time-average speed $v_{\rm ave}$.
%\textcolor{red}{
$\tilde{Q}$ seemed to increase in proportion to $v_{\rm ave}$.
%}
%\textcolor{red}{
When $v_{\rm ave}$ was sufficiently high,
a bacterial cell had the strong tendency to move counterclockwise along the brim of its pool.
This yielded a large value of $\tilde{Q}$.
When $v_{\rm ave}$ was low, on the other hand, 
we obtained a small value of $\tilde{Q}$ as shown in Fig.\ \ref{fig:ES_Q_properties}(b).
%}
From the distributions of cell speed $v$ for different values of $v_{\rm ave}$,
the distributions were found to have two peaks 
at $v\cong0$ \textmu m/s and $v>30$ \textmu m/s
as shown in 
Fig.\ \ref{fig:ES_speed_hist_and_trajectory}(a).
%One of them was almost at $v\cong0$ \textmu m/s and another one was at 
%\textcolor{red}{$v>30$} \textmu m/s.
%As $v_{\rm ave}$ increased, 
%the frequency of the peak at $v\cong0$ \textmu m/s became low, 
%while the frequency of the peak at $v\gtrsim30$ \textmu m/s became high.
%\textcolor{red}{
Under the condition that the total measuring times of $v$ is constant (18,000 successive cell speeds in time),
%}
%\textcolor{red}{
when the probability density at $v\cong0$ \textmu m/s was low,
that at 
$v>30$ \textmu m/s was high and then $v_{\rm ave}$ was large.
%}
%\textcolor{red}{
When the probability density at $v\cong0$ \textmu m/s was high, 
that at $v>30$ \textmu m/s was low and then $v_{\rm ave}$ was small.
These tendencies suggest that 
$v_{\rm ave}$ strongly depends on the relationship between the two peaks.
%}
%\textcolor{red}{
%Since each distribution was made from 18,000 successive speeds $v$ 
%and the total frequency was conserved, 
%it is considered to be effective that 
%we discuss the behavior of the distributions focusing on the two peaks approximately.
%In order to verify that $v_{\rm ave}$ is determined by the ratio of the frequencies at $v\cong0$ \textmu m/s
%to the frequencies at $v>30$ \textmu m/s,
In order to verify this dependence,
% of $v_{\rm ave}$, 
we simplified the calculation of the time-average speed $v_{\rm ave}$ 
by focusing on the two peaks in the distributions.
That is,
we introduced the simplified average speed $v^{\prime}_{\rm ave}$ defined by
%%%
\begin{equation}
v^{\prime}_{\rm ave} \
= \frac{v_{\rm s} p_{\rm s} + v_{\rm l} p_{\rm l}}{p_{\rm s} + p_{\rm l}}.
\label{eqn:vprime}
\end{equation}
%%%
%= v_{\rm s} \frac{p_{\rm s}}{p_{\rm s} + p_{\rm l}} + v_{\rm l} \frac{p_{\rm l}}{p_{\rm s} + p_{\rm l}} \
%
Here, $v_{\rm s}$ and $v_{\rm l}$ are the smaller and larger values of $v$, respectively.
$p_{\rm s}$ is the probability density at $v=v_{\rm s}$ 
and $p_{\rm l}$ is that at $v=v_{\rm l}$ (the inset in Fig. \ref{fig:Discussion_vprime_vave}).
%As shown in Fig.\ \ref{fig:Discussion_vprime_vave}(b), 
%the simplified average speed $v^{\prime}$ monotonically decreases as $p_{\rm s}$ increases.
%The data was characterized by the exponential function
%%%%
%\begin{equation}
%v^{\prime} = Ae^{\alpha p_{\rm s}},
%\label{eqn:exponential_function_vprime}
%\end{equation}
%%%%
%with the following fitting parameters
%%%%
%\begin{equation}
%A=49.2\ \text{\textmu m/s},\quad \alpha=-15.7.
%\label{eqn:fitting_parameters_vprime}
%\end{equation}
%%%%
Figure \ref{fig:Discussion_vprime_vave} shows the relationship between $v^{\prime}_{\rm ave}$ and $v_{\rm ave}$.
$v_{\rm ave}$ seemed to be linearly dependent on $v^{\prime}_{\rm ave}$, 
and to be approximately equal to $v^{\prime}_{\rm ave}$.
Thus, 
we can conclude that 
%we found that
$v_{\rm ave}$ is almost entirely determined by the behavior of the two peaks.
%This result is consistent with the $\lambda$ dependence of $v_{\rm ave}$ 
%as shown in Fig.\ \ref{fig:ES_speed_hist_and_trajectory}(c).
%} % end textcolor
%
%Therefore, we can say that 
%$v_{\rm ave}$ is determined by the ratio of the frequencies at $v\cong0$ \textmu m/s
%to the frequencies at 
%\textcolor{red}{$v>30$} \textmu m/s.
In the microscopic observation, 
a moving bacterial cell was observed to stop to change its direction.
Hence, the probability density 
%\textcolor{red}{$p_{\rm s}$}
%of the peak at $v\cong0$ \textmu m/s 
$p_{\rm s}$ represents the frequency of the change in the moving direction.
In other words, $v_{\rm ave}$ is considered to have a monotonically decreasing dependence 
on the frequency of the change in the moving direction.
Figure\ \ref{fig:ES_speed_hist_and_trajectory}(b) 
shows the trajectories of a bacterial cell for the different values of $v_{\rm ave}$.
There, 
we see that as $v_{\rm ave}$ increased, the trajectories became less complicated.
%\textcolor{red}{
%As shown in Fig.\ \ref{fig:ES_speed_hist_and_trajectory}(c), 
%$v_{\rm ave}$ did not depend on $\lambda$.}
When $v_{\rm ave}=54.3$ \textmu m/s, 
the trajectory was almost circular and $\tilde{Q}$ was 0.97.
In this case, 
the frequency of the change in the moving direction was almost zero and the distribution of $v$
had only one peak at $v\cong60$ \textmu m/s as shown in Fig.\ \ref{fig:Discussion_one_peak_histogram}.
%\textcolor{red}{
The peak at $v=v_{\rm s}$ disappeared from the distribution of $v$, 
whereas the peak at $v=v_{\rm l}$ remained in the distribution.
Hence, the value of $v_{\rm l}$ is considered to be the characteristic speed of a bacterial cell 
in a shallow circular pool.
%}
Furthermore, 
we observed that
a bacterial cell sometimes changed its moving direction at the same positions in a pool.
Some irregularities may exist at the bottom and the brim of a pool (the surface of an agar plate).
Note that the pool had a depth about 1 \textmu m,
which was the same scale as the diameter of a bacterial cell ($\simeq0.5$ \textmu m).
Therefore, 
we attributed the changes in the moving direction of a bacterial cell to the cell--noise interactions, 
%\textcolor{red}{
which were regarded as the interactions 
between each bacterial cell and the irregularities on the surface of a pool.
%}
From the above considerations, 
the behavior of a bacterial cell in the single cell systems
was suggested to be dependent on 
the competition between the cell--boundary interactions and the cell--noise interactions.

%%% 5節から.
From the above considerations,
we conclude the following.
%Let us summarize our findings.
In this experimental systems,
we were able to control the interaction range between bacterial cells by changing $\lambda$.
Such a situation is well described 
in Figs.\ \ref{fig:CM_relation_Q_lambda}(b) and \ref{fig:Discussion_Qt_R03}, 
which show the $\lambda$-dependence of the order parameter $Q$ defined in Sect.\ 3.1.1
and its modified version defined in Sect.\ 4.
The scattering of the plots of $Q$ in these graphs is caused by 
%\textcolor{red}{the effect of cell--noise interactions},
the effect of cell--noise interactions,
which were not controlled.
Nevertheless,
the parameter $\lambda$ is still relevant to describe the transitions,
since the critical value $\lambda_{\rm C1}$
seemed to be fixed at $\cong0.1$
so that the order parameter $Q$ evidently increased with $\lambda$ for 
%$\lambda\gtrsim0.1$,
%\textcolor{red}{$\lambda\geq0.1$}
$\lambda\geq0.1$
while the dependence of $Q$ on $\lambda$ was negligible for 
%$\lambda\lesssim0.1$ 
%\textcolor{red}{$\lambda<0.1$}
$\lambda<0.1$
for all systems that we examined.
Such a sharp transition at $\lambda=\lambda_{\rm C1}$ was not confirmed in the single cell systems.
Therefore, we conclude that 
in the transitions between the \textit{random motion} phase and the \textit{one-way rotational motion} phase,
the cell--cell interactions give the primary effect.

More detailed studies of the cell--noise interactions and of the cell--boundary interactions 
are required in order to give a more precise description of the transitions including critical phenomena.
In the present paper, 
we have reported our work on the transitions at $\lambda=\lambda_{\rm C1}$
with low values of $\rho$.
The transitions at $\lambda=\lambda_{\rm C1}$ with higher values of $\rho$ 
(between the \textit{turbulent motion} phase and the \textit{two-way rotational motion} phase)
as well as those at $\lambda=\lambda_{\rm C2}$ will be important future problems.
Extensions of the SPP model suitable to describe our experimental results are also required.

%\clearpage

%\clearpage

%%%%%%%%%%%%%%%%%%%%%%%%%%%%%%%%%%%%%%%%%%%%%%%%%%

%\clearpage
%\clearpage

%%%%%%%%%%%%%%%%%%%%%%%%%%%%%%%%%%%%%%%%%%%%%%%%%%
% Acknowledgements
%%%%%%%%%%%%%%%%%%%%%%%%%%%%%%%%%%%%%%%%%%%%%%%%%%
%\begin{acknowledgement}
\section*{Acknowledgements}
%\input{acknowledgement_v1}
%\input{acknowledgement_v2_20181212} % 20181212, katori check
%\input{acknowledgement_v3_20181213} % 20181213, katori check
%\input{acknowledgement_v4_20181215} % 20181215
%%%%%%%%%%%%%%%%%%%%%%%%%%%%%%%%%%%%%%%%%%%%%%%%%%

\hspace{15pt}
The authors would like to thank Mitsugu Matsushita, Helmut R. Brand, Takuma Narizuka, Yoshihiro Yamazaki, 
Yusuke T. Maeda, and Hirofumi Wada for useful discussion.
They also thank Makoto Katori for careful reading of the manuscript and useful comments.
JW is supported by a Chuo University Grant for Special Research 
and by a Grant-in-Aid for Exploratory Research (No.\ 15K13537)
from the Japan Society for the Promotion of Science (JSPS).
%%%%%%%%%%%%%%%%%%%%%%%%%%%%%%%%%%%%%%%%%%%%%%%%%%

%\newpage

%%%%%%%%%%%%%%%%%%%%%%%%%%%%%%%%%%%%%%%%%%%%%%%%%%
% bibliography
%%%%%%%%%%%%%%%%%%%%%%%%%%%%%%%%%%%%%%%%%%%%%%%%%%
%%% bibliography
%\input{bibliography_v1.tex}
%\input{bibliography_v2_20181213.tex}
%\input{bibliography_v3_20181215.tex}
%\input{bibliography_v4_20181216.tex}
%\input{bibliography_v5_20190208.tex}
%\input{bibliography_v6_20190318.tex}
%%%%%%%%%%%%%%%%%%%%%%%%%%%%%%%%%%%%%%%%%%%%%%%%%%

%%%%%%%%%%%%%%%%%%%%%%%%%%%%%%%%%%%%%%%%%%%%%%%%%%

%%% figs
%\input{figs_v1.tex}

\end{document}